\documentclass[aps,prr,groupedaddress,twocolumn]{revtex4-2}

\usepackage{Carrickc,lettrine}

\usepackage{enumitem}
\usepackage{mathtools, amsfonts, amssymb, amsthm, graphicx,xcolor,tikz,tkz-euclide, booktabs, tensor, empheq,subcaption,lipsum,float,subcaption}
\usepackage{siunitx}
\usepackage[capitalize]{cleveref}
\crefname{equation}{}{}

\newcommand{\eps}{\varepsilon}

\newcommand{\ovl}[1]{{\overline{#1}}}

\newcommand{\op}{\operatorname}

\usepackage{dsfont}

\newcommand{\nCa}{\mathit{Ca}}

\newcommand{\RR}{\mathbb{R}}
\renewcommand{\d}{\mathrm{d}}

\newtheorem*{claim*}{Claim}

\newtheorem*{conjecture*}{Conjecture}

\tikzstyle{reverseclip}=[insert path={(current page.north east) --
  (current page.south east) --
  (current page.south west) --
  (current page.north west) --
  (current page.north east)}
]

\begin{document}


\title{On the Geometry of Spreading Puddles}


\author{David Darrow}
\email[]{ddarrow@mit.edu}

\affiliation{Department of Mathematics, Massachusetts Institute of Technology, Cambridge, Massachusetts, U.S., 02139}


\date{\today}

\begin{abstract}
We develop a geometric model for the spreading of shallow, viscous puddles of arbitrary shape, building on the recent \emph{capillary current} model for axisymmetric droplets. In short, we assume that a spreading puddle remains close to instantaneous mechanical equilibrium as it spreads, with hydrostatic pressure balanced by surface curvature. In turn, its contact line advances so as to maximize the rate of energy loss subject to viscous dissipation. The resulting system yields a natural geometric evolution equation for both the two-dimensional footprint and the three-dimensional depth profile of a spreading puddle. In appropriate limits, it recovers the classical spreading laws for axisymmetric droplets, a local version of the Hoffman--Voinov--Tanner law for small non-axisymmetric puddles, and a nonlocal Hele-Shaw-like description for large, relatively regular puddles. We show that the model rationalizes new observations of silicone oil spreading over smooth borosilicate glass.
\end{abstract}

\ifx\wordct\undefined
\maketitle
\fi

\section*{Introduction}\label{sec:intro}
 We investigate the spreading of shallow, viscous puddles over flat surfaces. Specifically, we consider puddles with depth $h$ smaller than the capillary length $\ell_c=(\sigma/\rho g)^{1/2}$, where $\sigma$ is the surface tension of the liquid, $\rho$ is its density, and $g$ is the acceleration due to gravity. We also restrict attention to viscosity-driven puddles, i.e., those with characteristic spreading speed $U\lesssim U_0=\mu/\rho h$, where $\mu$ is the dynamic viscosity of the liquid. For water under normal ambient conditions, for instance, we consider puddles up to a maximum depth $\ell_c=2.7$~mm and a maximum spreading speed $U_0 =(\ell_c/h)\times 0.33$~mm~s$^{-1}$ (typically $1$--$10$~mm~s$^{-1}$).

Puddles of this sort are ubiquitous in daily life, in both the natural and built environment. Here, we focus primarily on the case of \emph{total wetting} of smooth surfaces (or equivalently, wetting of \emph{hydrophilic} surfaces), in which a puddle would eventually spread to a molecular-scale thickness if left alone. Total wetting occurs when machine oil spreads over smooth metal, when meltwater spreads over ice, and when rainwater spreads over certain plant surfaces~\citep{Bauer2009}. As we discuss later, the present work can be adapted straightforwardly to handle more general spreading processes, including total wetting of rough or porous surfaces (e.g., ink spills on paper) and partial wetting (e.g., water droplets on glass).


Much of the literature on puddle spreading has focused on the case of axisymmetric puddles, which we refer to as \emph{droplets}~\citep{Bonn2009,Popescu2012}. The behavior of a droplet depends greatly on the relative importance of gravity and surface tension, which in turn depends on how the horizontal droplet radius $R$ compares to the capillary length~$\ell_c$. Small droplets, with horizontal radius $R\ll \ell_c$, are well-described by the quasi-static model of \citet{Hervet1984}. Small droplets are entirely dominated by capillary effects. In short, they retain a spherical cap shape---always minimizing surface area---as they are spread outward by contact-line (i.e., droplet edge) forces. The spreading rate is determined by a global energy balance, recovering the classic Hoffman--Voinov--Tanner (HVT) laws \citep{Hoffman1975,Voinov1976,Tanner1979}
\begin{equation}\label{eq:HVT}
    \dot{R}\sim (\sigma/\mu)\theta^3,\qquad R\sim (\sigma/\mu)^{1/10}V^{3/10}t^{1/10},
\end{equation}
where $V$ is the total droplet volume, $\theta$ is the contact angle at the droplet edge, and $\mu$ is the dynamic viscosity of the liquid. \citeauthor{Hervet1984} also reported a logarithmic correction to \cref{eq:HVT}, later refined by \citet{Eggers2004}.

By contrast, \citet{Lopez1976} proposed that large droplets (for which $R\gg\ell_c$) exhibit a gravity-driven evolution, in which hydrostatic pressure gradients push the droplet outward. Their model yields a flattened pancake shape---as compared with the spherical cap shape adopted by small droplets---as well as the scaling
\begin{equation}\label{eq:1/8}
    R\sim (\rho g/\mu)^{1/8}V^{3/8}t^{1/8}.
\end{equation}
Notably, this scaling also arises in the theory of viscous gravity currents~\citep{Simpson1999,Huppert2006,Ungarish2009}. Viscous gravity currents are relatively deep ($h\gg\ell_c$) viscosity-driven flows, such as creeping lava flows or mudslides, for which surface tension effects are negligible; it may be seen as surprising, then, that they share the same spreading rate as large-but-shallow droplets. Subsequent authors have proposed various modifications to the large-droplet model of \citeauthor{Lopez1976}, including alternate descriptions of internal energy dissipation~\citep{Ehrhard1991,Ehrhard1993} and perturbative approaches to partially account for capillary effects~\citep{Voinov1995,Voinov1999}. Even still, the basic physical picture of gravity-driven spreading has remained widely accepted \citep{Bonn2009,Popescu2012}.

\citet{Darrow2026} recently introduced a unified \emph{capillary current} model for droplets of arbitrary size, extending the edge-driven model for small droplets and providing a consistent, edge-driven alternative to the gravity-driven model for large droplets. They posited that all viscous droplets maintain a quasi-equilibrium balance between hydrostatic and curvature pressure, perturbed only by interfacial forces at the contact line. The work performed by contact-line forces is balanced against global energy dissipation, which contains up to three components: viscous dissipation throughout the droplet bulk, viscous dissipation near the droplet edge, and microscopic friction immediately at the contact line. The capillary current model rationalizes new and old observations \citep{Darrow2026,Dorbolo2021} of \emph{Darcy precursor films}, i.e., thin films of fluid that percolate ahead of totally wetting droplets on rough substrates \citep{Cazabat1986,deGennes2003}. It also rationalizes previous observations for the small-to-large-droplet transition that occurs when the horizontal radius $R$ first exceeds $\ell_c$~\citep{Cazabat1986}, and it recovers previously-reported spreading rates of partially wetting droplets as they relax into equilibrium~\citep{deRuijter1999,Durian2022}. 

Prior work on non-axisymmetric puddle spreading is more limited, but several useful perspectives already exist. Weakly non-axisymmetric puddles can be described with perturbative methods~\citep{Greenspan_1978}, for instance, and more general puddles can be described with full-field equations, either via lubrication theory~\citep{Oron1997,Schwartz1998} or diffuse-interface methods~\citep{Jacqmin2000}. Closest in spirit to the present work is the mathematical literature on quasi-static droplets, in which an instantaneously surface-area-minimizing geometry is coupled to a local law for contact-line motion~\citep{Alberti2011,Feldman2013}. This literature offers substantial insight into the mathematical properties of small puddles in absence of gravity, and we will see that it arises in one limit of the present work. Finally, we note that substantial work has been done to understand the local structure of a moving contact line \citep{Huh1971,hocking_spreading_1982,Cox1986,deGennes1990,Snoeijer2013}. We expect that much of this literature applies directly to non-axisymmetric puddles, but we do not revisit this local theory here. Rather, we take the perspective that, regardless of the global puddle geometry, the fine structure of the droplet edge depends only on the local contact angle $\theta$.

In the present work, we develop a generalized capillary current model to account for spreading puddles of arbitrary shape.
Specifically, if a puddle of volume $V$ initially covers a region $\Omega_0\subset\RR^2$ of a flat, hydrophilic substrate, what shape does the puddle take, and how does this shape evolve over time? We deduce a geometric evolution equation that predicts the puddle's 2-D footprint $\Omega_t$ and 3-D depth profile $h(\vec{x},t)$ over time. The recovered model yields a relatively simple, geometric picture for spreading puddles, and serves to unify several distinct spreading regimes. In generalizing the work of \citet{Darrow2026}, it automatically recovers the HVT laws \cref{eq:HVT} in the limit of small, axisymmetric droplets, as well as the widely-observed scaling \cref{eq:1/8} for large, axisymmetric droplets.  For small puddles of arbitrary shape, our model yields a local form of the HVT law, which falls into the class of surface-area-minimizing droplet models from the mathematical literature~\citep{Alberti2011,Feldman2013}. For large, relatively regular puddles, it yields a nonlocal spreading law reminiscent of classical Hele-Shaw flow \citep{Hele-Shaw1898,Gustafsson2006}; Hele-Shaw flow describes the motion of fluid between two closely-spaced parallel plates, and provides a prototype for moving-boundary phenomena more generally.

After deriving the generalized capillary current model and discussing its various limits, we report new experimental results of silicone oil spreading over borosilicate glass. We observe a close match to the predictions of the present model, by comparing to a numerical simulation of the same. We then outline how---up to a reparameterization of time, and supposing contact-line friction is negligible---capillary current spreading can be seen as a purely geometric flow, independent of liquid volume and material parameters. Finally, we discuss how the present work might be leveraged to study a broader class of spreading problems.

\section*{The capillary current model}
We first derive our generalized capillary current model based on the macroscopic fluid mechanics of non-axisymmetric puddles. We assume here that the substrate is smooth, uniform, and hydrophilic (i.e., the system undergoes total wetting), and that the liquid is non-volatile, meaning that it does not evaporate or undergo chemical reactions. Much attention has been given to more general spreading problems, involving partial wetting \citep{deGennes1990,BrochartWyart1992,Petrov1992,deRuijter1997,deRuijter1999,Eggers2005,Durian2022,Butt2022}, rough or non-uniform substrates \citep{Cazabat1986,Starov2002,Starov2002b,Starov2003,Grzelakowski2009,Gambaryan2014,Dorbolo2021,Butt2022}, volatile liquids \citep{Goncalves2022,Wang2024,Lee2025}, and deep, porous substrates \citep{Frank2012,Gambaryan2014,Chebbi2021}. The extension of the present work to partial wetting and to rough substrates is straightforward, and the axisymmetric case of each is discussed in \citet{Darrow2026}. Volatile liquids and deep, porous substrates are beyond the present scope, but we expect the present model to provide a strong foundation for further investigation.

Below, we fix length, time, and mass scales such that
\begin{equation}\label{eq:nondim}
    \sigma = \rho g = \mu = 1.
\end{equation}
In particular, lengths are scaled by the capillary length $\ell_c = (\sigma/\rho g)^{1/2}$ and velocities are scaled by the `visco-capillary speed' $u_c =\sigma/\mu$. Note that the latter is distinct from (and often much larger than) the maximum speed $U_0=\mu/\rho h$ of viscosity-driven spreading; consequently, the non-dimensionalized spreading speed (or equivalently, the capillary number $\nCa = U/u_c$) is often quite small for shallow, viscous puddles. 

Our generalized capillary current model is founded on two hypotheses:
\begin{enumerate}
    \item A viscous puddle is always shaped so as to minimize its instantaneous configuration energy.
    \item A viscous puddle spreads so as to maximize energy decay subject to dissipative forces.
\end{enumerate}
We investigate these hypotheses in turn.

\begin{figure}
    \centering
    \includegraphics[width=1\linewidth]{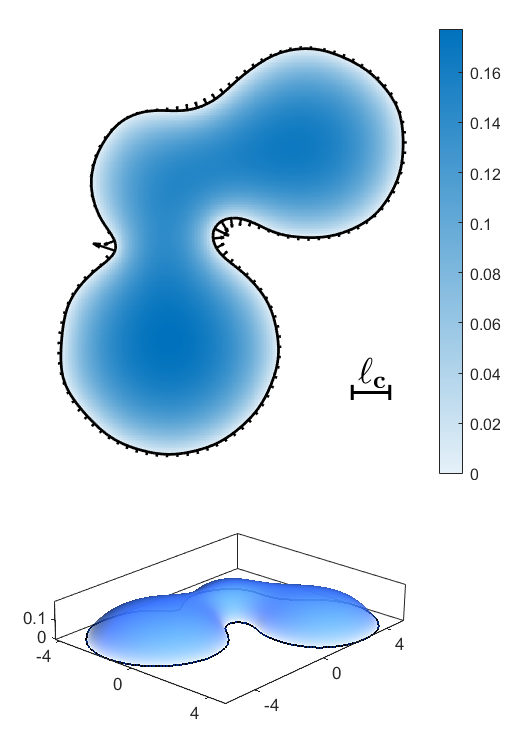}
    \caption{The capillary current model reproduces the qualitative features one expects of a non-axisymmetric puddle, such as a relatively flat depth profile, a boundary layer of thickness $\ell\lesssim\ell_c$, and a spreading velocity that tends to smooth out the contact line. We depict the calculated depth profile and spreading velocity of a three-lobed puddle, which we return to in our experimental results. The colorbar and 3-D rendering are presented in units of $\ell_c$, and the latter is exaggerated in the vertical direction for visual clarity.}
    \label{fig:profile}
\end{figure}

First, to derive the instantaneous 3-D shape of a puddle, we note that the configuration energy is composed of (i) gravitational potential energy and (ii) interfacial energy. Noting our non-dimensionalization \cref{eq:nondim}, the gravitational potential energy takes the form
\[E_\mathrm{grav} = \tfrac{1}{2}\int_{\Omega_t} h(\cdot,t)^2\,\d A,\]
where $h(\vec{x},t)$ is the depth profile of the puddle and $\Omega_t\subset\RR^2$ is its 2-D footprint. For a liquid totally wetting a smooth substrate, the puddle is generally preceded by a microscopic \emph{precursor film} of fluid \citep{Hardy1919,Popescu2012}, so it effectively spreads over a pre-wet substrate. It is thus convenient to offset the interfacial energy of the puddle-substrate system by that of a uniform, pre-wet substrate. After doing so, we find that the interfacial energy of the system is proportional to the difference between the surface area of the puddle and the 2-D area $A(t)$ of the footprint $\Omega_t$, so we find
\[E_\mathrm{surf} = \int_{\Omega_t}\sqrt{1+|\nabla h(\cdot,t)|^2}\,\d A - A(t) \approx \tfrac{1}{2}\int_{\Omega_t}|\nabla h|^2\,\d A,\]
written to leading order in $h/\ell_c$ and $|\nabla h|$. The case of partially wetting liquids is similar, except that one would have to account more carefully for the energy of the liquid-substrate interface. In any case, the total energy takes the form
\begin{equation}\label{eq:energy}
    E(t) = \int_\Omega\left(\tfrac{1}{2} h(\cdot,t)^2 + \tfrac{1}{2}|\nabla h(\cdot,t)|^2\right)\d A,
\end{equation}
and the energy-minimizing depth profile (with fixed volume $V$ and boundary $\partial\Omega_t$) is the solution of the following Helmholtz equation:
\begin{equation}\label{eq:profile}
    h(\vec{x},t) - \nabla^2 h(\vec{x},t) = h_0(t),\qquad h|_{\partial\Omega_t} = 0,
    \end{equation}
    where $h_0(t)$ is fixed by the integral constraint 
    \begin{equation}\label{eq:h_int}
        \int_{\Omega_t} h(\cdot,t)\,\d A = V.
    \end{equation}
Recalling our non-dimensionalization \cref{eq:nondim}, one can see that the equation \cref{eq:profile} expresses a balance between the hydrostatic pressure $p_\mathrm{grav}=\rho g h$ and the curvature pressure $p_\mathrm{curv} = -\sigma\nabla^2h$. The latter interpretation reveals that $h_0(t)$ is the pressure at the puddle-substrate interface; under our quasi-equilibrium assumption, this pressure is spatially constant. \Cref{fig:profile} depicts the calculated depth profile of a three-lobed puddle, which we will return to in our experimental results.

\begin{figure*}
    \centering
    \begin{subfigure}{\textwidth}
    \includegraphics[width=\linewidth]{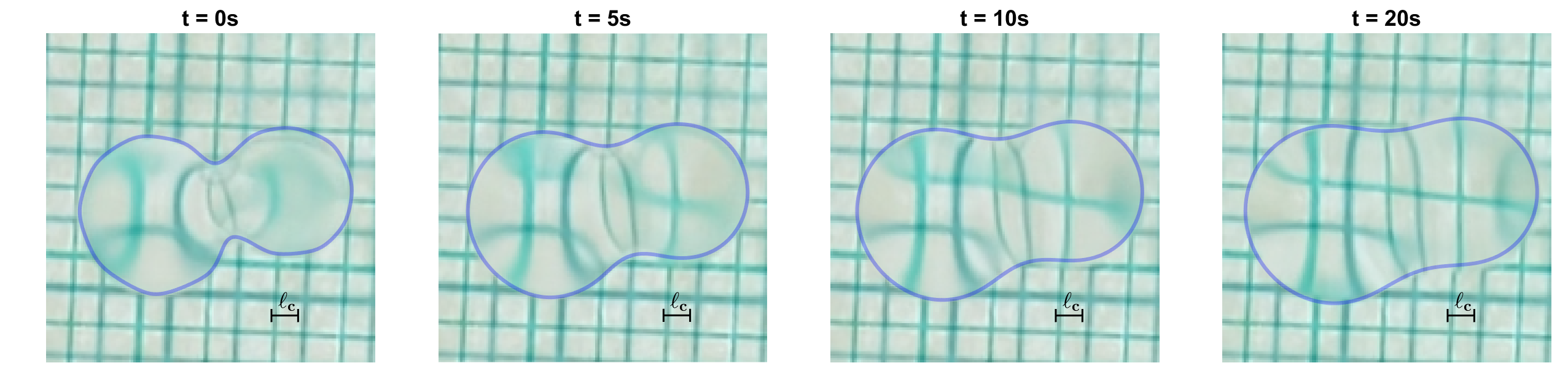}
    \caption{Dumbbell-shaped puddle, $V=20$~\SI{}{\micro\liter}}
    \end{subfigure}
    \begin{subfigure}{\textwidth}
    \includegraphics[width=\linewidth]{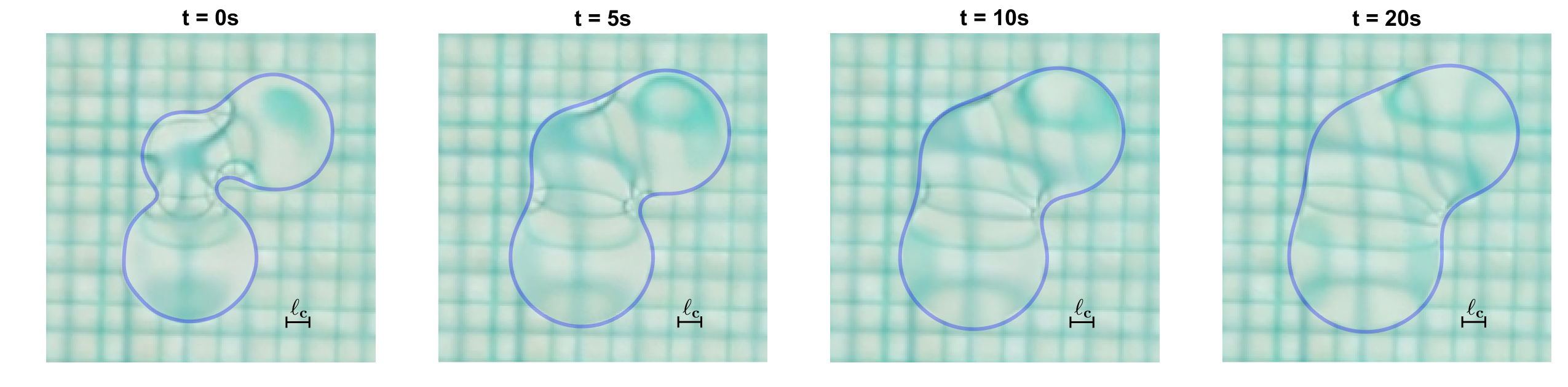}
    \caption{Three-lobed puddle, $V=30$~\SI{}{\micro\liter}}    
    \end{subfigure}
    \caption{The extended capillary current model rationalizes new observations of silicone oil spreading over smooth borosilicate glass. We depict the spreading of \textbf{(a)} a dumbbell-shaped puddle of $50$~cSt silicone oil with volume $V=20$~\SI{}{\micro\liter} and \textbf{(b)} a three-lobed puddle with volume $V=30$~\SI{}{\micro\liter}, both compared against simulations of the capillary current model (blue curves) with $\alpha=120$ and $\beta=\kappa=0$. The initial state of each simulation is manually traced to match the puddle geometry at $t=0$~s, with the aid of a cubic B\'ezier spline (Appendix~\ref{app:simulation}). We observe a satisfactory match to experiment in both cases.}
    \label{fig:main}
\end{figure*}

We now consider our second hypothesis, that a capillary current spreads so as to shed its configuration energy as quickly as possible. By appealing to Young's law \citep{Young1832}---or by taking the `shape derivative' of the energy (see Appendix~\ref{sec:shape})---one can calculate the time derivative of $E(t)$ explicitly:
\begin{equation}\label{eq:Edot}
    \dot{E}(t) = -\tfrac{1}{2}\int_{\partial\Omega_t} \theta(\cdot,t)^2\,u(\cdot,t)\,\d\ell,
\end{equation}
where $\theta = -\partial_{\hat{n}}h$ is the local contact angle and $u$ is the (\emph{a priori} unknown) normal velocity along the contact line $\partial\Omega_t$. We note that the formula \cref{eq:Edot} applies only to totally wetting liquids; the formula for partial wetting is similar, but depends on the substrate-dependent {equilibrium contact angle} $\theta_\mathrm{eq}>0$. 

Quantitatively, we claim that the spreading speed $u$ should minimize the integral \cref{eq:Edot}---it remains only to identify appropriate constraints for this minimization problem. In fact, just as in the axisymmetric case, the spreading rate is constrained by the rate of energy dissipation in the puddle, which we denote by $D(t)$. Typically, one would hope to calculate $D(t)$ according to its standard integral representation:
\begin{equation}\label{eq:D_naive}
    D(t)\stackrel{?}{=} \int_{\Omega_t}\int_0^{h(\cdot,t)} \mu |\nabla\vec{u}_\mathrm{3D}(\cdot,z,t)|^2\,\d z\,\d A,
\end{equation}
where $\vec{u}_\mathrm{3D}$ is the 3-D velocity field within the puddle. Unfortunately, the gradient $\nabla\vec{u}_\mathrm{3D}$ is known to blow up at the boundary, causing the integral \cref{eq:D_naive} to diverge \citep{Huh1971}.

One way to proceed is by truncating the integral \cref{eq:D_naive} a microscopic distance away from the boundary, similar to the treatment of small, axisymmetric droplets by \citet{Hervet1984}. We believe this strategy may be worth pursuing, but we here develop a physically-motivated alternative. Following \citet{Darrow2026}, we decompose the energy dissipation into three components: viscous dissipation throughout the droplet bulk, viscous dissipation within a distance $\ell\lesssim\ell_c$ of the droplet edge, and microscopic friction immediately at the contact line:
\begin{equation}\label{eq:intro_diss}
    D = \alpha D_\mathrm{bulk} + \beta D_\mathrm{edge} + \kappa D_\mathrm{micro},
\end{equation}
with parameters $\alpha,\beta,\kappa\geq 0$ quantifying their relative magnitudes. Heuristically, we expect $\alpha D_\mathrm{bulk}$ to be dominant for relatively large puddles, $\beta D_\mathrm{edge}$ to be dominant for relatively small puddles, and $\kappa D_\mathrm{micro}$ to be relatively small for all totally wetting liquids.

The microscopic and edge-localized dissipation can be calculated using local versions of the standard arguments (presented in Appendix \ref{app:axi}), yielding
\[D_\mathrm{micro} \sim \int_{\partial\Omega_t} u^2\,\d\ell,\qquad D_\mathrm{edge}\sim \int_{\partial\Omega_t} u^2/\theta\,\d\ell.\]
We note that a more careful treatment of the edge-localized viscous dissipation would yield the logarithmic correction identified by \citet{Hervet1984} and \citet{Eggers2004}.

The bulk dissipation is somewhat more involved. For this, we suppose that, away from the boundary $\partial\Omega_t$, the puddle is approximately flat with depth $h_0(t)$. This assumption is expected to be most accurate for relatively large puddles, whose depth profile varies weakly away from an edge region of thickness $\ell\lesssim\ell_c$. For smaller puddles, edge-localized dissipation dominates, and the particular form of $D_\mathrm{bulk}$ is less consequential.

The resulting system can be described using a standard lubrication approximation \citep{Batchelor1967}.  In short, we suppose that the fluid velocity adheres to a Poiseuille-type profile:
\begin{equation}\label{eq:poiseuille}
    \vec{u}_\mathrm{3D}(\vec{x},z,t) = \frac{3}{2}\,\frac{z(2h_0(t)-z)}{h_0(t)^2}\,\vec{u}_\mathrm{2D}(\vec{x},t),
\end{equation}
and that the fluid admits a 2-D velocity potential $\phi$ such that $\vec{u}_\mathrm{2D} = \nabla\phi$. The continuity equation then yields
\begin{equation*}
    \nabla^2\phi(\vec{x},t) = -\dot{h}_0(t)/h_0(t),\qquad \partial_{\hat{n}}\phi|_{\partial\Omega_t} = u.
\end{equation*}
Integrating over $\Omega_t$ and comparing with Green's theorem allows us to remove the dependence on $\dot{h}_0$:
\begin{equation}\label{eq:pot}
    \nabla^2\phi(\vec{x},t) = \frac{1}{A(t)}\int_{\partial\Omega_t}u(\cdot,t)\,\d\ell,\qquad\partial_{\hat{n}}\phi|_{\partial\Omega_t} = u,
\end{equation}
again writing $A(t)$ for the 2-D area of $\Omega_t$.

Plugging the expression \cref{eq:poiseuille} into the formula \cref{eq:D_naive} for viscous dissipation and integrating by parts shows that 
\begin{equation}\label{eq:bulk_closure}
    D_\mathrm{bulk} = \int_{\Omega_t} \int_0^{h_0(t)}|\partial_z\vec{u}_\mathrm{3D}|^2\,\d z\,\d A\sim\int_{\partial\Omega_t}\phi u/h_0\,\d \ell,
\end{equation}
where, without loss of generality, we offset the velocity potential by a constant to ensure that
\begin{equation}\label{eq:phi_int}
    \int_{\Omega_t}\phi(\cdot,t)\,\d A= 0.
\end{equation}

Finally, comparing the dissipation rate to the known rate of work \cref{eq:Edot}, we arrive at the total energy balance
\begin{equation}\label{eq:totalbalance}
    \int_{\partial\Omega_t}\theta^2u\,\d\ell = \int_{\partial\Omega_t}\left[\alpha \phi u/h_0 + \beta u^2/\theta + \kappa u^2\right]\,\d\ell,
\end{equation}
folding the $\frac{1}{2}$ factor of $\dot{E}(t)$ into the dissipation coefficients for convenience. Minimizing the integral \cref{eq:Edot} subject to the constraint \cref{eq:totalbalance} yields the following Euler--Lagrange equation:
\begin{equation}\label{eq:eullag}
    \theta(\vec{x},t)^2 = \alpha \phi(\vec{x},t)/h_0(t) + \beta u(\vec{x},t)/\theta(\vec{x},t) + \kappa u(\vec{x},t).
\end{equation}
This equation is linear in $u(\vec{x},t)$, since $\phi$ is; inverting it allows one to recover $u$ from the depth profile $h$. An example velocity field is depicted alongside the three-lobed puddle in \Cref{fig:profile}, with parameters $\alpha = 120$ and $\beta=\kappa=0$. Together, the equations \cref{eq:profile}, \cref{eq:pot}, and \cref{eq:eullag} and the integral constraints \cref{eq:h_int} and \cref{eq:phi_int} yield a closed system for the evolution of $\Omega_t$ and $h(\vec{x},t)$.


\section*{Limits of interest}
Several special cases of the capillary current model bear mentioning. In the axisymmetric case, the present model reduces to that of \citet{Darrow2026}, up to appropriate modifications of $\alpha$, $\beta$, and $\kappa$, so it produces the correct predictions for axisymmetric droplets of all sizes (Appendix~\ref{app:axi}). In particular, it yields the HVT laws \cref{eq:HVT} for small axisymmetric droplets ($\op{diam}\Omega_t\ll\ell_c$), the scaling \cref{eq:1/8} for large axisymmetric droplets ($\op{diam}\Omega_t\gg\ell_c$), and the appropriate transitionary regime for droplets of intermediate size~\citep{Cazabat1986}. 

We also note that the present model places 1-D and 2-D droplet spreading on the same footing. Recall that the work of \citet{Darrow2026} was able to offer a separate derivation for 1-D capillary currents, thereby recovering the $R\sim t^{1/7}$ scaling reported by \citet{Tanner1979} and \citet{Mchale1995} for thin stripes of fluid. Here, we see how the separate 1-D and 2-D capillary current models presented by \citeauthor{Darrow2026} are simply limiting cases of a more general, geometric model.

Next, for small, totally wetting puddles of arbitrary shape, the edge-localized dissipation should dominate, and we recover a local version of the HVT law:
\[u(\vec{x},t) = \beta^{-1}\theta(\vec{x},t)^3.\]
As noted above, a more careful treatment of the edge-localized dissipation would reveal a logarithmic correction to this local HVT law~\citep{Hervet1984,Eggers2004}. Separately, for relatively small, partially wetting puddles, the contact-line friction may dominate~\citep{deRuijter1999,Durian2022}, in which case we recover the relation
\[u(\vec{x},t) = \kappa^{-1}\theta(\vec{x},t)^2.\]
In both cases, if one neglects the influence of gravity on the puddle geometry---for instance, if the puddle is sufficiently small---the evolution falls into a class of local, quasi-static flows studied in the mathematical literature~\citep{Alberti2011,Feldman2013}.

At the other extreme, for large puddles with horizontal radius of curvature everywhere large (compared to $\ell_c$), one expects the bulk dissipation to dominate, and for the contact angle $\theta=\theta(t)$ to be approximately spatially uniform. Inverting \cref{eq:eullag} in this limit, we see that the spreading speed is given by $u=\partial_{\hat{n}}\phi|_{\partial\Omega_t}$, where $\phi$ solves the following problem:
\begin{equation}\label{eq:heleshaw}
    \nabla^2\phi(\vec{x},t) = \mathrm{const.},\qquad \phi|_{\partial\Omega_t} = \alpha^{-1}h_0(t)\theta(t)^2,
\end{equation}
with the constraint 
\[\int_{\Omega_t}\phi(\cdot,t)\,\d A=0.\]
This problem can be seen as a variant of classical Hele-Shaw flow \citep{Hele-Shaw1898,Gustafsson2006} with uniform fluid injection over $\Omega_t$ and modulating ambient pressure.

\section*{Experimental results} We compare the predictions of the capillary current model to experimental observations of 50~cSt silicone oil spreading on smooth borosilicate glass. The silicone oil we use (63148-62-9, Millipore Sigma) is non-volatile~\citep{ecetoc2011} and totally wets borosilicate glass~\citep{Dorbolo2021}, so it fits the assumptions of our capillary current model. It has a capillary length of $\ell_c=1.5$~mm and a visco-capillary speed of $u_c=0.44$~m~s$^{-1}$, so we expect it to be more gravity-dependent and to spread slower than a water puddle (for which $\ell_c=2.7$~mm, $u_c=81.8$~m~s$^{-1}$) of the same size.


For each of our experiments, we place a $75$~mm$\times75$~mm square of glass above a backlit surface. We place gridded paper under the glass, such that the contact line can be detected via optical distortion of the grid. We fasten a 50~MP camera overhead, recording at 1 frame per second. Finally, we place multiple small (10~\SI{}{\micro\liter}) droplets of silicone oil on top of the glass, which quickly coalesce into a single, irregular puddle, and we observe the spreading of this puddle starting 5~s after the final droplet is placed.

\begin{figure}
    \centering
    \includegraphics[width=\linewidth]{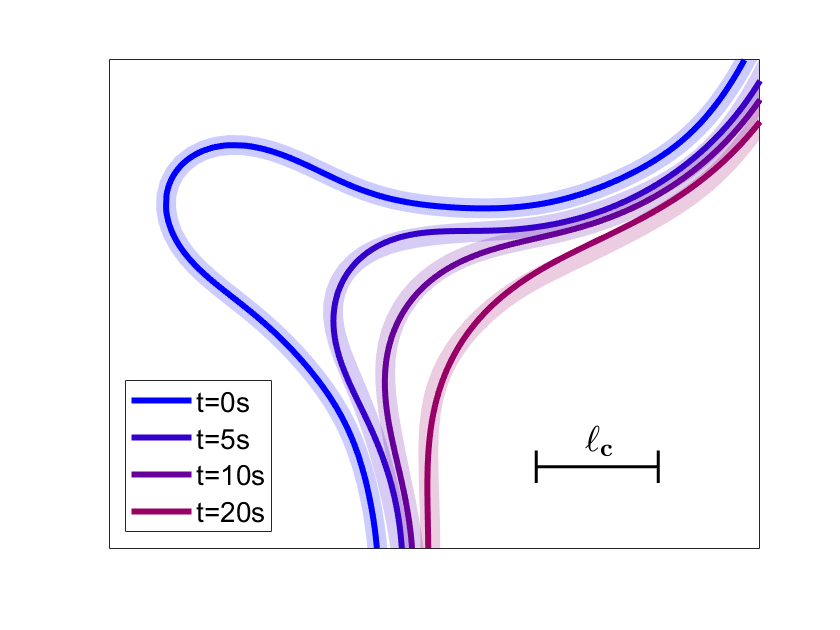}
    \caption{Spreading of the concave region in the three-lobed puddle (\Cref{fig:main}\emph{b}). Solid curves represent simulation results, while the thicker shaded curves correspond to the true (manually traced) puddle boundary at each timestamp, $\pm 1$~pixel ($0.12$~mm).}
    \label{fig:zoom}
\end{figure}

In the first experiment, the liquid initially forms a dumbbell shape of volume $V=20$~\SI{}{\micro\liter} and maximum diameter $12$~mm (\Cref{fig:main}\emph{a}). In the second, the liquid initially forms a three-lobed shape of volume $V=30$~\SI{}{\micro\liter} and maximum diameter $15$~mm (\Cref{fig:main}\emph{b}), corresponding to the 3-D geometry shown in \Cref{fig:profile}. Each puddle slowly spreads outward until reaching a more regular oval shape. We depict $20$~s of this spreading process in \Cref{fig:main}\emph{a} and \Cref{fig:main}\emph{b}, respectively, and compare against simulations of the capillary current model (blue curves) with $\beta=\kappa=0$ and a single fit parameter $\alpha=120$; because the puddles are relatively large compared to the capillary length, we expect bulk dissipation to dominate. Details of the numerical simulation are given in Appendix~\ref{app:simulation}. In both experiments, we observe a satisfactory match between the visible edge of the puddle and the simulated contact line over the simulated time horizon. \Cref{fig:zoom} shows the concave region of the three-lobed puddle in greater detail. The solid curves represent simulation results at $t=0,5,10,20$~s, respectively, and the thicker shaded curves represent the manually-traced puddle boundary at the same timestamps (Appendix~\ref{app:simulation}), with error bars of $\pm 1$~pixel ($0.12$~mm).


\section*{Capillary Currents as a Geometric Flow}
Finally, we discuss how to reparameterize the time coordinate such that the evolution of $\Omega_t$ can be determined entirely by its instantaneous 2-D geometry. This reparameterization offers a valuable new perspective on capillary current spreading, clarifying how it relates to classical geometric flows (such as Hele-Shaw flow \citep{Hele-Shaw1898,Gustafsson2006}) and isolating its dependence on physical parameters.

As a first step, we fix $\kappa=0$; since we expect microscopic friction to be dominated by viscous dissipation for totally wetting liquids, this is a relatively mild assumption. We note that the maximum pressure $h_0(t)$ inside a capillary current is decreasing over time. From \cref{eq:profile}, we see that the characteristic contact angle $\theta$ tends to decrease over time as $\theta\sim h_0(t)$, and thus, from \cref{eq:eullag}, that the characteristic spreading speed $U$ tends to decrease over time as $U\sim h_0(t)^3$. To counteract this effect, we define the reparameterized time coordinate
\[\tau(t) = \int^t h_0(t')^3\,\d t',\]
corresponding to the scaled spreading velocity 
\[v(\vec{x},\tau) = u(\vec{x},\tau)/h_0(\tau)^3.\]

The depth field $h(\vec{x},\tau)$ must be rescaled likewise; inspecting \cref{eq:profile}, we see that the rescaled field $\eta(\vec{x},\tau) = h(\vec{x},\tau)/h_0(\tau)$ satisfies a depth-independent Helmholtz equation
\begin{equation}\label{eq:profile_rescaled}
    \eta(\vec{x},\tau) - \nabla^2\eta(\vec{x},\tau) = 1,\qquad \eta|_{\partial\Omega_\tau} = 0,
\end{equation}
and yields a rescaled contact angle 
\[\zeta(\vec{x},\tau) = -\partial_{\hat{n}}\eta(\vec{x},\tau)|_{\partial\Omega_\tau} = \theta(\vec{x},\tau)/h_0(\tau).\]
In these coordinates, the Euler--Lagrange equation \cref{eq:eullag} becomes
\begin{equation}\label{eq:eullag_rescaled}
    \zeta(\vec{x},\tau)^2 = \alpha\hat{\phi}(\vec{x},\tau) + \beta v(\vec{x},\tau)/\zeta(\vec{x},\tau),
\end{equation}
with the rescaled velocity potential $\hat{\phi}$ defined by
\begin{equation}\label{eq:pot_rescaled}
    \nabla^2 \hat{\phi}(\vec{x},\tau) = \mathrm{const}.,\qquad \partial_{\hat{n}}\hat{\phi}|_{\partial\Omega_\tau}=v,\qquad \int_{\Omega_\tau}\hat{\phi}\,\d A=0.
\end{equation}
The equations \cref{eq:profile_rescaled,eq:eullag_rescaled,eq:pot_rescaled} form a closed system for the evolution of $\Omega_\tau$, independent of the puddle volume $V$, pressure $h_0(t)$, and material parameters $\sigma$, $\rho$, and $\mu$. 

The geometric flow we have recovered is dynamically interesting in its own right. For one, while the original capillary current system yields the modulating Hele-Shaw-like flow \cref{eq:heleshaw} for large, relatively regular puddles, the reparameterized system converges to a true Hele-Shaw flow in the same limit. It also offers concrete predictions for viscous puddle spreading. For instance, it suggests that---for a fixed initial contact line $\partial\Omega_0$---a deeper puddle (still satisfying $h\lesssim\ell_c$) will spread \emph{faster} than a shallower puddle, but both will pass through the same sequence of 2-D geometries. Moreover, we now see that the timescale of spreading always scales as $\tau_{\mathrm{spreading}}\sim V^{-3}$, generalizing the results for small droplets \cref{eq:HVT} and large droplets \cref{eq:1/8}.

\section*{Discussion and perspectives}
We have proposed a generalized capillary current model for the spreading of shallow, viscous puddles of arbitrary shape. Our model reproduces known spreading laws for axisymmetric droplets and thin stripes of fluid, and it rationalizes new observations of non-axisymmetric oil puddle spreading on glass. Moreover, it shows how---up to a reparameterization of time---the 2-D footprint of capillary currents can be understood as a fully geometric flow, independent of puddle volume or composition.

The present work suggests several avenues for further study. In one direction, it might be worthwhile to partially lift the assumptions of our model. The quasi-equilibrium profile \cref{eq:profile} assumes a relatively shallow puddle ($h\lesssim\ell_c$) and relatively shallow depth gradients ($|\nabla h|\lesssim 1$), for instance, but both assumptions could be lifted at the cost of introducing a nonlinear equation for $h(\vec{x},t)$. In another direction, the closure \cref{eq:bulk_closure} we have used to define the bulk viscous dissipation $D_\mathrm{bulk}$ is only approximate, especially for puddles with strongly irregular interiors. As discussed above, one could circumvent this limitation by calculating the total dissipation $D(t)$ via a truncation of the full integral \cref{eq:D_naive}.

Finally, much recent work has attempted to understand relatively exotic limits of droplet spreading, including liquid evaporation \citep{Goncalves2022,Wang2024,Lee2025}, liquid imbibition into porous media \citep{Gambaryan2014,Dorbolo2021, Chebbi2021}, and contact angle hysteresis \citep{Johnson1964} in partial wetting scenarios~\citep{Butt2022}. In revisiting the basic physics of viscous puddle spreading, the present work offers a flexible foundation on which to develop more general models. As an example, we remark that a critical experimental observation of \citet{Darrow2026} was the spreading rate of a {Darcy precursor film} ahead of a totally wetting droplet on a rough substrate \citep{Cazabat1986,deGennes2003}. Following arguments similar to those presented here, one could develop a model for Darcy precursor films ahead of non-axisymmetric puddles.


\ifx\wordct\undefined

\section*{Acknowledgments}
We would like to thank Lucas Warwaruk (MIT Mechanical Engineering) and John W.~M.~Bush (MIT Mathematics) for their support, and for providing assistance with experiments. We would also like to acknowledge the support of an NDSEG Graduate Fellowship.


\appendix

\setcounter{secnumdepth}{2}

\section{Axisymmetric capillary currents}\label{app:axi}
We review the physics of axisymmetric droplet spreading, as described by the capillary current model of \citet{Darrow2026}. Here, the 2-D footprint $\Omega_t$ of the droplet is a disk of radius $R(t)$, and the dynamics depend only on the radial coordinate $r=|\vec{x}|$. The Helmholtz equation \cref{eq:profile} can be solved in closed-form:
    \begin{equation}\label{eq:profile_axi}
    \begin{gathered}
        h(r,t) = h_0(t)\left[1 - \frac{I_0(r)}{I_0(R(t))}\right],\\ h_0(t) = \frac{V/\pi}{R(t)^2 - 2 R(t) \frac{I_1(R(t))}{I_0(R(t))}},
        \end{gathered}
    \end{equation}
where $I_0(r)$ denotes the modified Bessel function of the first kind \citep{Abramowitz1965}. For large droplets ($R\gg \ell_c$), this profile yields a flat, pancake-like shape of depth $h_0$, which drops to zero within a distance $\sim\ell_c$ of the contact line. For small droplets ($R\ll \ell_c$), this profile yields a spherical cap with radius of curvature $2\ell_c^2/h_0$. In both cases, the droplet-substrate interface has constant pressure $\rho g h_0$. 

Now, the spreading velocity $u(\vec{x},t)$ is itself axisymmetric, so one can deduce it uniquely from the energy balance 
\[R\theta^2 u = \alpha D_\mathrm{bulk} + \beta D_\mathrm{edge} + \kappa D_\mathrm{micro}\]
once we find appropriate expressions for the dissipation rates $D_\mathrm{bulk}$, $D_\mathrm{edge}$, and $D_\mathrm{micro}$.

First, the value of $D_\mathrm{micro}$ depends only on the spreading speed and the arclength of the contact line, and not on the contact angle. Buckingham's theorem \citep{Birkhoff2015} yields
\[D_\mathrm{micro} \sim \mu Ru^2,\]
consistent with models of partial wetting \citep{deRuijter1999,Durian2022}. 

Next, \citet{Hervet1984} use cutoff lengths to estimate the viscous dissipation near the edge of the droplet, i.e., within a distance $\ell\lesssim\ell_c$ of the contact line. They model this region as a triangular wedge with contact angle $\theta$, and the velocity profile as a vertical shear with fixed speed $u$ along the droplet surface:
\[\vec{u}_\mathrm{3D} \sim \frac{uz}{\theta(R-r)}\hat{r},\qquad |\nabla\vec{u}_\mathrm{3D}|\approx |\partial_z\vec{u}_\mathrm{3D}| \sim \frac{u}{\theta(R-r)}.\]
They introduce a microscopic cutoff scale $a\ll\ell_c$, and integrate the dissipation over $r\in(R-\ell,R-a)$ to find
\[D_\mathrm{edge} \sim \mu R\log(\ell/a) \,u^2/\theta.\]

Finally, \citet{Darrow2026} estimate the viscous dissipation in the droplet bulk using a scaling argument, which we can simplify as follows:
\begin{equation}\label{eq:bulk_axi}
    D_\mathrm{bulk}\sim \mu Vu^2/h^2\sim\mu R^4u^2/V,
\end{equation}
noting from \cref{eq:profile_axi} that $V\sim hR^2$ holds for both large and small droplets (albeit, with slightly different coefficients). 

Setting $\mu=1$ and $\dot{R} = u$, one thus finds the following evolution equation for $R$:
\[\dot{R} = \frac{R\theta^2}{\alpha R^4/V + \beta R/\theta + \kappa R},\]
writing $\theta =\theta(R,V) = -\partial_rh|_{R}$ for the contact angle. For small droplets, the capillary current model reduces to that of \citet{Hervet1984}, and one recovers the well-established HVT laws \cref{eq:HVT}. For large droplets, the capillary current model yields a self-consistent, edge-driven explanation for the scaling \cref{eq:1/8} first reported by \citet{Lopez1976}. The capillary current model also rationalizes new and old observations of (Darcy) precursor films on rough substrates \citep{Darrow2026,Dorbolo2021}, as well as observations for the small-to-large-droplet transition that occurs when $R\sim\ell_c$~\citep{Cazabat1986}. Finally, after adapting the capillary current model to the case of partial wetting~\citep{Darrow2026}, it offers an edge-driven explanation for the predictions of \citet{deRuijter1999} and \citet{Durian2022}. 


\section{Shape derivative of the energy}\label{sec:shape}
We here derive the formula \cref{eq:Edot} for the time derivative of the energy, by taking the `shape derivative' of the energy functional \citep{Delfour2011}. One can see this argument as a derivation of Young's law \citep{Young1832} for totally wetting puddles of arbitrary shape. Indeed, Young's law states that the interfacial force per unit arclength along the contact line is $f=\frac{1}{2}\sigma\theta^2$, yielding exactly the formula \cref{eq:Edot} for the total work applied by interfacial forces.

To proceed, we suppose the puddle boundary has a prescribed outward normal speed $u(\vec{x},t)$, and study how the configuration energy $E(t)$ varies over time. Applying Hadamard's formula for the shape derivative \citep{Delfour2011}, one finds the time derivative of the energy to be
\begin{align*}
    \dot{E}(t) &= \int_{\Omega_t} \left(\nabla h(\cdot,t)\cdot\nabla\partial_th(\cdot,t) + h(\cdot,t)\partial_th(\cdot,t)\right)\d A \\
    &\qquad + \tfrac{1}{2}\int_{\partial\Omega_t} \left( |\nabla h(\cdot,t)|^2 + h(\cdot,t)^2\right)u(\cdot,t)\,\d\ell.
\end{align*}
This expression can be simplified greatly. First, noting that the height $h(\vec{x},t)$ vanishes on $\partial\Omega_t$ allows us to remove one of the two boundary terms. Subsequently integrating the volume integral by parts and applying the Helmholtz equation \cref{eq:profile} yields
\begin{align*}
    \dot{E}(t)&= h_0(t)\int_{\Omega_t} \partial_th(\cdot,t)\,\d A + \int_{\partial\Omega_t} \partial_{\hat{n}}h(\cdot,t)\,\partial_t h(\cdot,t)\,\d\ell\\
    &\qquad + \tfrac{1}{2}\int_{\partial\Omega_t} |\nabla h(\cdot,t)|^2\,u(\cdot,t)\,\d\ell.
\end{align*}
Volume conservation implies that the first integral must vanish. Furthermore, we know that
\begin{equation*}
    \partial_t h(\vec{x},t)=-u(\vec{x},t)\partial_{\hat{n}} h(\vec{x},t)\qquad \text{for}\;\vec{x}\in\partial\Omega_t.
\end{equation*}
Putting these pieces together, we arrive at the desired equation \cref{eq:Edot}.

\section{Numerical simulation of the model}\label{app:simulation}
The simulation results presented in \Cref{fig:main,fig:zoom} consist of two primary steps: post-processing experimental data to estimate the contact line geometry at $t=0$, and integrating this geometry forward under the capillary current model to obtain predictions for later times.

For post-processing, we first perform a linear transformation of pixel data to map the boundary of our $75$~mm$\times75$~mm glass panel to a square of side length~$50$, ensuring that the capillary length $\ell_c=1.5$~mm is set to unity. In both experiments, we fix the initial simulation time $t=0$ to be approximately $5$~s after the final droplet is placed. To trace the initial contact line, we mark at least 30 points on the boundary of each puddle ($\pm 1$~pixel), connect these points with a cubic B\'ezier spline~\citep{Mortenson1999}, and resample with 1200 equispaced points $z_j$.

The integration step proceeds as follows. Given a dense sampling of points $z_j\in\partial\Omega_t\subset\mathbb{C}$, we first approximate the area $A(t)$ of $\Omega_t$ 
by calculating the following contour integral using a midpoint method:
\[A(t) = \frac{1}{2i}\oint_{\partial\Omega_t} \ovl{z}\,\d z.\]
We then smooth $\partial\Omega_t$ with eight passes of a periodic binomial filter to recover smoothed sample points $\tilde{z}_j$, and calculate an approximate outward normal vector $\hat{n}_j\propto -i(\tilde{z}_{j+1}-\tilde{z}_{j-1})$ at each smoothed point. We construct a new, `outer' curve $\Gamma$ that connects the points
\[\tilde{w}_j = \tilde{z}_j + \eps\hat{n}_j,\qquad \eps = 0.2\sqrt{A(t)/\pi}\]
with a cubic spline, and resample it with 300 equispaced points $w_j\in\Gamma$. Next, we find the least-squares solution $(c_1,...,c_{300})$ to
\[F(z_k)=\sum\nolimits_j c_jK_0(z_k-w_j) \approx 1,\qquad \forall z_k,\]
where $K_0$ denotes a modified Bessel function of the second kind~\citep{Abramowitz1965}. The function $\eta = 1 - F$ approximately solves the rescaled equation \cref{eq:profile_rescaled}, and we can estimate the rescaled volume under $\eta$ by integrating \cref{eq:profile_rescaled} over $\Omega_t$,
\[\tilde{V}=\int_{\Omega_t}\eta\,\d A = A + \int_{\partial\Omega_t}\partial_{\hat{n}}\eta\,\d\ell,\]
and computing the latter integral using a midpoint method. We rescale $h=(V/\tilde{V})\eta$ to compute the true depth profile, and compute $h_0(t)$ and $\theta(z_k,t)$ directly.

Next, we compute $\phi$ via a similar least-squares problem, choosing $a$ and $(b_1,...,b_{300})$ in the expression
\[\phi(z) = a|\vec{x}|^2 + \sum\nolimits_j b_k\log|z-w_j|\]
to minimize the squared residual in \cref{eq:eullag} over all $z_k$. We subtract the mean value $A(t)^{-1}\int\phi\,\d A$ directly; the latter can be converted to a line integral over $\partial\Omega_t$ using Green's theorem and calculated via a midpoint method. 

Finally, we calculate $u(z_j,t)$ as the normal derivative of $\phi$, and step $z_j \mapsto z_j + \delta u(z_j,t)$. We perform a mild implicit Laplace smoothing to the resulting curve, i.e., we map $\vec{z} \mapsto (1 - \ell^2\nabla^2)\vec{z}$, with smoothing parameter $\ell = 0.35\times \op{mean}|z_{k+1}-z_k|\approx 0.01$. This final step ensures that small artifacts do not develop over long integration horizons, but does not substantially change the short-term evolution. In both simulations, both least-squares problems are satisfied with residual always $<10^{-4}$.

\setcounter{secnumdepth}{-1}

\bibliography{foo}

\fi

\end{document}